\documentclass[11pt,a4paper]{article}
\usepackage[all]{xy}
\usepackage{amsmath,amsfonts,amssymb,amsthm,amscd,pb-diagram,pb-xy}
\usepackage{graphicx}
\usepackage{epsfig}
\usepackage{caption}

\usepackage{color}  

\definecolor{red-}{rgb}{0.8,0.0,0.0}

\definecolor{green-}{rgb}{0.0,0.7,0.0}

\definecolor{blu-}{rgb}{0.0,0.0,1.0}

\newenvironment{Remark}{\par \medskip \noindent{\sc Remark:}}

\newcommand{\Euclideo}{\ensuremath{\mathbb{E}}}
\def\vth{\vartheta}

\def\A{\mathcal A\/}
\def\B{\mathcal B\/}
\def\C{\mathcal C\/}

\def\F{\mathcal F\/}

\def\M{\mathcal M\/}

\def\S{\mathcal S\/}

\def\={\, = \,}
\def\de#1/de#2{\frac{\partial {#1}}{\partial {#2}}}
\def\De#1/de#2{\dfrac{\partial {#1}}{\partial {#2}}}

\def\vett#1{\underline{\mathbf {#1}}}

\begin{document}

\title{Notes on a paper by F. G{\'e}not and B. Brogliato on the Painlev{\'e} paradox}

\author{Stefano Pasquero\\
	Departments of Mathematics, Physics and Computer Sciences \\
	University of Parma \\
	Parco Area delle Scienze 53/a, 43124 PARMA -- Italy \\
	E-mail: stefano.pasquero@unipr.it \\
	ORCID: https://orcid.org/0000-0001-9261-8542
}
\maketitle
\date
\begin{abstract}
	\noindent
	We reconsider the analysis of the Classical Painlev\'e Problem developed by F.~G{\'e}not and B.~Brogliato (\cite{GenBro} {\em New results on Painlev{\'e} paradoxes.} -
	{\em European Journal of Mechanics-A/Solids}, 18(4):653--677, 1999), focusing on the consistency of their results with Galilean invariance. We show that certain conclusions concerning the  dynamical evolution of the mechanical system are not invariant under changes of Galilean observer and therefore cannot, in their present form, be interpreted as intrinsic properties of the system. In particular, we show that the classification of motion states depends on the observer through the velocity--dependent characterization of the frictional constraint, and we trace the origin of this dependence to the Galilean velocity-addition theorem. We present and discuss possible reformulations of the model aimed at restoring observer--invariant descriptions of the problem.
\end{abstract}

\newpage

\section*{Introduction}
The Painlevé paradox, first described by Jellett (\cite{Jellet}) in 1872 and subsequently studied and formalized by Painlevé (\cite{Painleve}), from whom it takes its name, consists in a violation of the determinism principle of Classical Mechanics. This phenomenon arises in certain mechanical systems within the framework of rigid-body theory and subject to unilateral rough constraints.

\begin{figure}[h!]
	\begin{center}
		\includegraphics[width=0.8\textwidth]{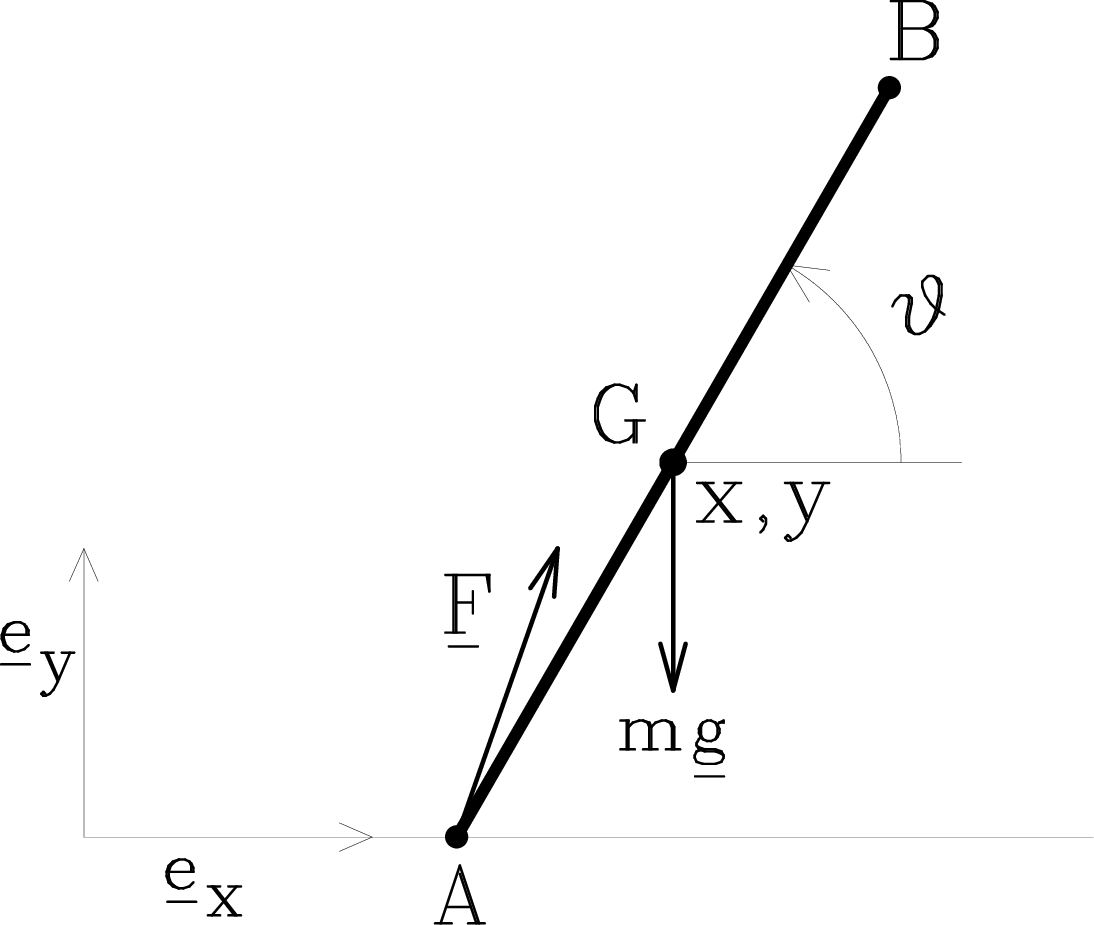}
	\end{center}
	\caption{\label{Fig1} The Classical Painlev\'e Problem}
\end{figure}

The best-known of these mechanical systems, whose study is commonly referred to as the Classical Painlevé Problem (hereafter CPP), consists of a homogeneous rigid rod $AB$ of length $2L$ and mass $m$, moving in the upper vertical half-plane bounded by a rough horizontal line satisfying the Amontons–Coulomb–Morin  dry-friction laws (hereafter ACM). The rod is subjected only to its weight $m \vett{g}$ and to the possible contact reaction $\vett{F}$ exerted by the line. The contact is assumed to occur only at the endpoint $A$ of the rod (see Figure \ref{Fig1}).

Within the vast literature on the subject, the highly cited 1999 paper by Frank Génot and Bernard Brogliato (hereafter GB) \cite{GenBro} plays a significant role, both because of the results obtained therein and because of the influence that its methodology and analysis have had on subsequent studies of the CPP and on the related literature. In their paper, GB exhibit the paradoxical nature of the CPP by showing that, for sufficiently large values of the friction coefficient $\mu$ of the supporting line, there exist motions of the system for which a (smooth, in the sense of being described by sufficiently regular functions) dynamical evolution either does not exist or is not unique. They then analyze the motion of the system in contact with the rough line, qualitatively determining its possible evolution.

\smallskip

In the present note,  we aim to point out some aspects of the GB analysis that appear problematic from the standpoint of Galilean invariance. We show that some of the conclusions drawn by GB regarding the behaviour of the mechanical system depend on the inertial observer analyzing the CPP, and that a different observer—even within the class of inertial observer—following the same methodology and analytical procedure, may infer qualitatively different behaviours.

\medskip

In Section 1, we briefly summarize the methodology adopted by GB and present their first significant results. In Section 2, we point out the aforementioned critical issues in GB’s analysis when varying the Galilean observer studying the CPP. In Section 3, we explain the origin of these issues, essentially on the basis of the Galilean velocity-addition theorem and the definition of sliding velocity. We also show that certain features of the results obtained by GB ought to remain invariant under a change of observer, even beyond the class of inertial observers. In Section 4, we propose a substantial and generally applicable modification of the constraint description, supplementing the usual geometric characterization with a kinematic description of the points of the constraint. Such a modification is necessary, in particular, for rough constraints when the mechanical problem is analyzed through Newton’s second law. In Section 5, we present a reformulation of the procedure followed by GB, based on the kinematic description of the constraint and consistent with the aforementioned invariance requirements. In particular, we show that, depending on the type of rough constraint considered, this reformulation may be immediately effective, effective only after substantial adjustments, or altogether inapplicable. In the latter case, it becomes evident that an approach to the CPP based on the standard form of Newton’s second law is not especially suitable, whereas an approach based on Newton’s second law formulated for impulsive systems is preferable.

The bibliography has deliberately been kept to the minimum necessary to keep the paper self-contained. A substantial portion of the literature on the CPP can be found, in addition to the aforementioned paper \cite{GenBro}, in the review article \cite{ChampVar2016}.

\section{Summary of the GB Paper}

In \cite{GenBro}, GB analyze the CPP by means of Newton’s second law and the dry-friction laws ACM, investigating their compatibility with the contact condition. In particular (see Figure \ref{Fig1}), they formulate the CPP in the vertical half--plane $\Euclideo_2^+$ by introducing a (not specified in detail) Galilean observer  with coordinates $(x,y,\vth)$,  where $x,y$ denote the Cartesian coordinates of the rod’s center of mass $G$ (with $y \geq 0$), and $\vartheta \in (0,\pi)$ denotes the rod orientation with respect to the horizontal. Relative to this observer, they introduce the unilateral rough constraint expressed by the condition $y_A\ge0$ and they write the cardinal equations of Dynamics (i.e., the balance equations of linear and angular momentum), which follow directly from Newton's second law, projected onto the natural horizontal, vertical, and out-of-plane directions. Setting (with obvious notation) $\vett{F} \= F_T \, \vett{e}_x + F_N \, \vett{e}_y$ and denoting by $I \= \frac13 m L^2$ the moment of inertia of the rod, they obtain
\begin{eqnarray}\label{EqCard}
	\left\{
	\begin{array}{lcl}
		m \ddot{x} &\=& F_T \\
		m \ddot{y} &\=& -mg + F_N \\
		I \ddot{\vth} &\=& L \left(-F_N \cos\vth + F_T \sin \vth\right)
	\end{array}
	\right.
\end{eqnarray}
These equations are complemented by the laws of dry friction. Denoting by $\vett{v} \= v \, \vett{e}_x$ the sliding velocity of the rod along the line at point $A$, they assume
\begin{eqnarray}\label{DryFrict}
	\left\{
	\begin{array}{lclcl}
		|F_T| &\, \le \,& {\phantom -}\mu \, |F_N|  &\quad {\textrm{if}} \quad & \vett{v} \= 0 \\ \\
		F_T &\, = \,& - \, \mu \, |F_N| \dfrac{v}{|v|} &\quad {\textrm{if}} \quad & \vett{v} \ne 0
	\end{array}
	\right.
\end{eqnarray}

GB then classify the motion of the system into four possible modes. The mode $M_I$ occurs when the rod is not in contact with the line. In this case, the reaction exerted by the line vanishes, i.e., $F_T \=  F_N \= 0 $, and the evolution of the system is straightforward.

Assuming instead that the contact condition $y_A \= y - L \sin\vth \=0$ between the endpoint $A$ of the rod and the line is satisfied, three further modes arise. Observing that $\dot{x}_A \= \dot{x} \, + \, L \dot{\vth} \sin\vth$, one has:
\begin{description}
	\item[mode $M_{II}$:]  $\dot{x}_A < 0$, from which (\ref{DryFrict}) yields $F_N  \ge 0 $ and $F_T \= \mu  F_N $;
	\item[mode $M_{III}$:]  $\dot{x}_A > 0$, from which (\ref{DryFrict}) yields $F_N  \ge 0 $ and $F_T \= - \mu  F_N $;
	\item[mode $M_{IV}$:]  $\dot{x}_A = 0$, from which (\ref{DryFrict}) yields $F_N  \ge 0 $ and $- \mu F_N \le F_T \le \mu  F_N $.
\end{description}
It is worth noting already at this stage that the first and most significant critical aspect of the GB analysis arises precisely in this classification, where the assumption $v \= \dot{x}_A$ is implicitly adopted.

The authors then observe that mode $M_{III}$ is equivalent to mode $M_{II}$ upon selecting a different Galilean observer, and therefore restrict the analysis to the latter case. Since in this situation
$F_T \= \mu  F_N$, substitution into (\ref{EqCard}) yields:
\begin{eqnarray}\label{EqMoto}
	\left\{
	\begin{array}{lcl}
		m \ddot{x} &\=& \mu  F_N \\
		m \ddot{y} &\=& -mg + F_N \\
		I \ddot{\vth} &\=& L  \left(- \cos\vth + \mu \sin \vth\right)  F_N
	\end{array}
	\right.
\end{eqnarray}

To complete the framework of the analysis, GB introduce the conditions:
\begin{eqnarray}\label{LCP}
	\ddot{y}_A \ge 0; \quad F_N \ge 0; \quad \ddot{y}_A F_N \= 0.
\end{eqnarray}

\medskip

The analysis proceeds by differentiating twice the contact condition $y_A \= y - L \sin\vth \=0$, and then, using (\ref{EqMoto}) together with $I\= \frac13 m L^2$, expressing $\ddot{y}_A$ in the form $\ddot{y}_A \= - \A + \B \, F_N$, where
\begin{eqnarray}\label{CoeffAB}
	\A(\vth,\dot{\vth}) \= g - L \dot{\vth}^2\sin\vth ; \qquad \B(\vth,\mu) \= \dfrac{1}{m} \left(1+3\cos\vth(\cos\vth - \mu \sin\vth)\right)
\end{eqnarray}
A first set of results is obtained by comparing these coefficients with conditions (\ref{LCP}):
\begin{itemize}
	\item if $\A > 0, \B >0$, there exists a unique value $F_N \= \dfrac{\A}{\B}$ satisfying conditions (\ref{LCP});
	\item if $\A < 0, \B >0$, there exists a unique value $F_N \= 0$ satisfying conditions (\ref{LCP});
	\item if $\A > 0, \B <0$, no value of $F_N$ satisfies conditions (\ref{LCP}) (a smooth temporal evolution of the Painlev\'e system does not exist);
	\item  if $\A < 0, \B <0$, two values of $F_N$, namely $F_N \= 0 $ and $F_N \= \dfrac{\A}{\B}$, satisfy conditions (\ref{LCP}) (the smooth temporal evolution is not unique).
\end{itemize}
In particular, the system is deterministic if and only if $\B >0$, whereas it is non-deterministic (in two distinct ways depending on the sign of $\A$) when $\B <0$. The argument is completed by showing that there exist values of the friction coefficient $\mu$, namely $\mu> \frac43$, for which there exist values of $\vth$ such that the coefficient B becomes negative, thereby establishing the existence of the paradox.

\bigskip

\begin{figure}[h!]
	\begin{center}
		\includegraphics[width=1\textwidth]{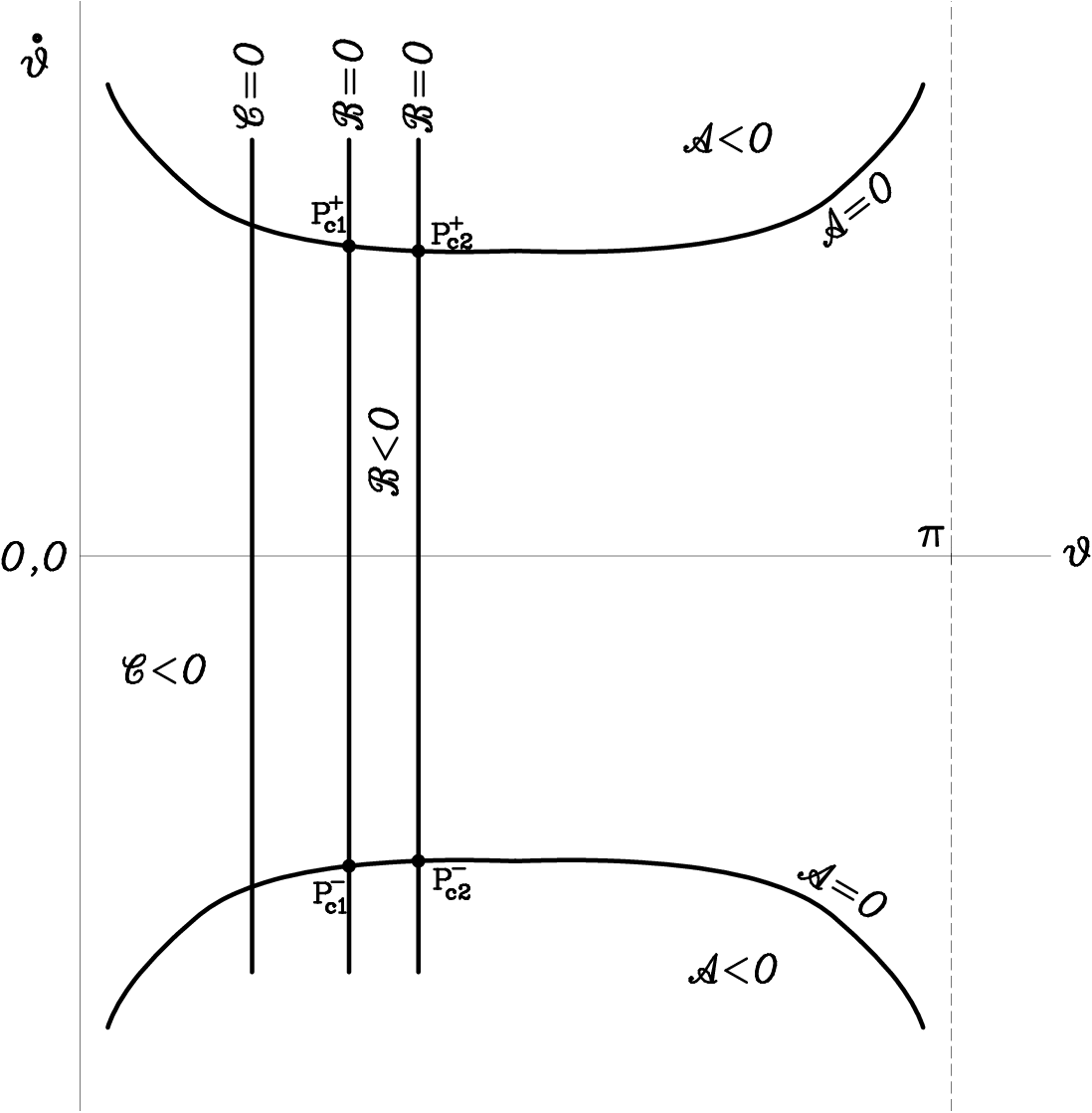}
	\end{center}
	\caption{\label{Fig2} Qualitative phase-space diagram of the CPP motion state}
\end{figure}

The qualitative analysis is further refined by evaluating, again from (\ref{EqMoto}), the angular acceleration $\ddot{\vth}$ of the rod as a function of the instantaneous velocity field of the rod and of the component $F_N$ of the reaction, whenever the latter exists. From the third equation in (\ref{EqMoto}), it follows immediately that, if $F_N\=0$, then $\ddot{\vth} \=0$. Otherwise, if $F_N \= \dfrac{\A}{\B}$,  the angular acceleration is governed by the differential equation
\begin{eqnarray}\label{EqAccAng}
	\ddot{\vth}(\vth,\dot{\vth},\mu) \= \C(\vth,\mu) \dfrac{\A(\vth,\dot{\vth})}{\B(\vth,\mu)}
\end{eqnarray}
where $\C \= - \frac{3}{mL}(\cos\vth - \mu \sin\vth)$. For $\mu>\frac43$, a sign analysis of the functions $\A, \B, \C$ in the $(\vth,\dot{\vth})$--plane, reveals the existence of distinguished points $P_{c1}^-, P_{c1}^+, P_{c2}^-, P_{c2}^+$, corresponding to intersections of the zero level sets of these functions (see the qualitative phase-space diagram reported on page 3 of \cite{GenBro}, briefly illustrated in Figure \ref{Fig2}).

A second set of results is obtained by analyzing the behaviour of the system, evolving along an arbitrary trajectory in the $(\vth,\dot{\vth})$-space, as it approaches these critical points. The analysis shows that, in some cases, detachment of the rod from the line occurs without the contact point coming to rest; in other cases, the contact point comes to rest and detachment follows; and in yet other cases, an “impact without collision” occurs, giving rise to an impulsive behaviour characterized by a jump discontinuity in the velocities $\Delta \vett{v}_A \= \vett{v}^R_A - \vett{v}^L_A$, so that the evolution of the system exists, but it is not smooth.

\section{Critical Issues in the GB Analysis}

The fundamental premise of this section concerns the invariance of the laws of dynamics with respect to the choice of observer within the class of Galilean observers. It is an immediate consequence of Galileo’s principle and of the very definition of a Galilean observer that all Galilean observers measure the same forces and the same accelerations for a mechanical system in motion. Therefore, all dynamical properties of a mechanical system--namely, those derived from the application of Newton's second law--must hold for every Galilean observer. Consequently, if a property established for a mechanical system by one inertial observer does not hold for other inertial observers, then such a property must be regarded as observer-dependent rather than intrinsic to the system.

With this observation in mind, we now show that the results correctly derived by GB in (\cite{GenBro}) hold for certain inertial observers but not for all inertial observers.

Let  $\F$ denote the inertial observer chosen by GB to study the motion of the rod, and let $(x,y,\vth)$ be the corresponding Lagrangian coordinates. Consider the observer $\F^*$
defined by the Lagrangian coordinates $(x^*,y^*,\vth^*)$ given by
\begin{eqnarray}\label{NuoveCoord}
	\left\{\begin{array}{lcl}
		{x^*} &=&  {x} + u \, t \\
		{y^*} &=& {y} \\
		{\vth^*} &=& {\vth}
	\end{array} \right.
\end{eqnarray}
where the constant velocity $u$ is arbitrarily chosen. Clearly, $\F^*$, being in uniform translational motion with respect to $\F$, is itself an inertial observer, and one has
\begin{eqnarray}\label{LeggeVarVelAcc}
	\left\{\begin{array}{lcl}
		\dot{x}^* &=&  \dot{x} + u \\
		\dot{y}^* &=& \dot{y} \\
		\dot{\vth}^* &=& \dot{\vth}
	\end{array} \right. \qquad
	\left\{\begin{array}{lcl}
		\ddot{x}^* &=&  \ddot{x}  \\
		\ddot{y}^* &=& \ddot{y} \\
		\ddot{\vth}^* &=& \ddot{\vth}
	\end{array} \right.
\end{eqnarray}
The observers $\F$ and $\F^*$
therefore measure the same vertical positions and orientations, hence the same vertical velocities and accelerations and the same angular velocities and accelerations, as well as the same horizontal accelerations, but different horizontal velocities. In what follows, we denote by an asterisk only those quantities relative to $\F^*$ that differ from the corresponding ones in $\F$; for instance, the horizontal component of the velocity of point $A$ in $\F^*$
is denoted by
$\dot{x}_A^*$, whereas the vertical component is still denoted by $\dot{y}_A$.

It follows immediately from (\ref{LeggeVarVelAcc}) that, for any values of $\dot{x}, \vth, \dot{\vth}$
such that
$\dot{x}_A \= \dot{x} \, + \, L \dot{\vth} \sin\vth <0$, the arbitrariness of $u$ implies that there exist values of $u$ such that
\begin{eqnarray}\label{VelocitaNuova}
	\dot{x}^*_A \= \dot{x}^* \, + \, L \dot{\vth} \sin\vth \=  \dot{x} \, + \, u \, + \, L \dot{\vth} \sin\vth \, > 0 \, .
\end{eqnarray}
This implies that there exist inertial observers for which, while for $\F$ one has $F_T \= \mu  F_N $ and the corresponding balance equations are those given in (\ref{EqCard}), for $\F^*$ one has instead $F^*_T \= - \mu  F_N$ (which is already unacceptable in itself, since the forces acting on the rod must be the same for both $\F$ and $\F^*$), and the corresponding equations become
\begin{eqnarray}\label{EqMotoAsterisco}
	\left\{
	\begin{array}{lcl}
		m \ddot{x}^* &\=& - \mu  F_N \\
		m \ddot{y} &\=& -mg + F_N \\
		I \ddot{\vth} &\=& - L  \left(\cos\vth + \mu \sin \vth\right)  F_N
	\end{array}
	\right.
\end{eqnarray}

It is straightforward to verify that conditions (\ref{LCP}) for $\F$ and $\F^*$
coincide, as do the expressions for $\A$ and $\A^*$. This is not the case, however, for $\B$ and $\B^*$, since
\begin{eqnarray}
	\B^* \= \dfrac{1}{m} \left(1+3\cos\vth(\cos\vth + \mu \sin\vth)\right) \ne \B
\end{eqnarray}

Repeating for $\F^*$
the same analysis carried out by GB, one finds that the Painlev\'e paradox manifests itself for $\F^*$
under the same types of conditions as for $\F$: it does not occur if $\B^* >0$; the time evolution of the system does not exist if $\A > 0 , \B^* <0$; and it is not unique if $\A < 0 , \B^* < 0$. Moreover, also for $\F^*$, the term $\B^*$ can be negative only if $\mu > 4/3$.

However, for any fixed value of $\mu$, the functions $\B_{\mu}(\vth)$ and $\B_{\mu}^*(\vth)$ are symmetric with respect to the axis $\vth \= \frac{\pi}{2}$. This implies that, if $\mu > \frac43$, there exist values of $\overline{\vth}$ such that $\B(\overline{\vth}) < 0$ while $\B^*(\overline{\vth}) > 0$ (and, by symmetry between the observers $\F, \F^*$, also the converse occurs).

This reveals a first problematic aspect in the results of GB: the Painlev\'e paradox may arise for one inertial observer and not arise for another inertial observer. We will return later to the existence of, and the suitable choice of, the inertial frame to be adopted for the frame invariant description of the CPP.

\bigskip

\begin{figure}[h!]
	\begin{center}
		\includegraphics[width=0.7\textwidth]{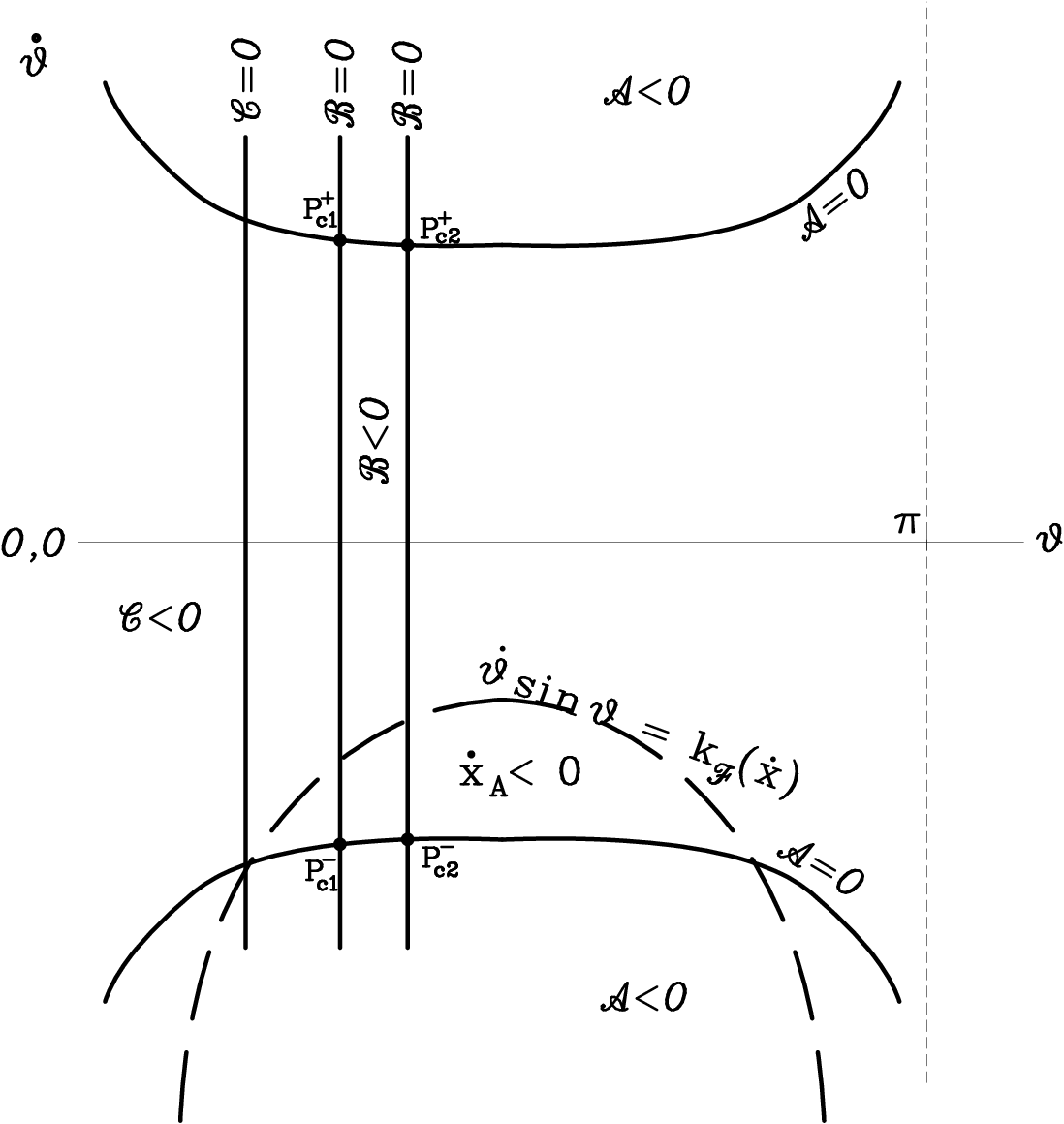}
	\end{center}
	\caption{\label{Fig3} Phase-space diagram including  $\dot{x}_A<0,  k_{\F}(\dot{x})<0$}
\end{figure}

\begin{figure}[h!]
	\begin{center}
		\includegraphics[width=0.7\textwidth]{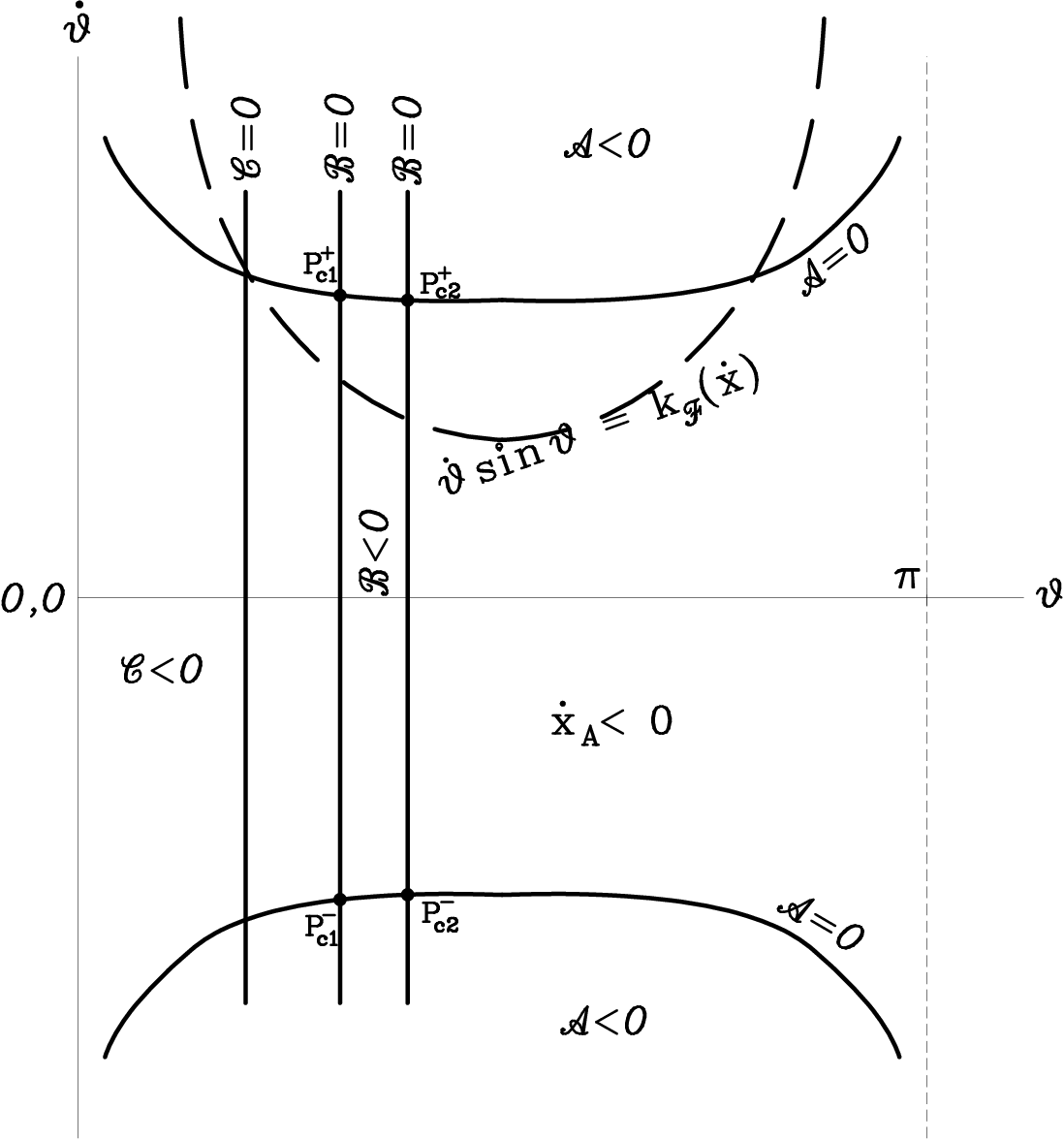}
	\end{center}
	\caption{\label{Fig4} Phase-space diagram including  $\dot{x}_A<0,  k_{\F}(\dot{x})>0$}
\end{figure}

A second (and in fact double) problematic aspect emerges from the analysis of the diagram associated with the differential equation (\ref{EqAccAng}), qualitatively reproduced in Fig. 2. That diagram is derived, for the observer $\F$, under the assumption that the condition $\dot{x}_A \= \dot{x} \, + \, L \dot{\vth} \sin\vth <0$ holds. The condition
\begin{eqnarray}
	\dot{\vth} \sin\vth < - \dfrac{\dot{x}}{L} \= k_{\F}(\dot{x})
\end{eqnarray}
must therefore be incorporated into the phase-space analysis (see Figures \ref{Fig3}, \ref{Fig4}).
Then, an issue arises from the fact that, even for the single observer $\F$, depending on the sign and the value of
$\dot{x}$ and then of $k_{\F}(\dot{x})$, some, or possibly all, of the critical points $P_{c1}^-, P_{c2}^-,  P_{c1}^+, P_{c2}^+$	are excluded from the phase--plane analysis. This may actually be viewed as an improvement over the results presented by GB, as it refines the analysis by showing that the dynamical behaviour of the system may depend not only on the inclination of the rod and its angular velocity, but also on its linear velocity.

Unfortunately, taking into account that the value of $\dot{x}$ is independent of $\vth$ and $\dot{\vth}$, and that, under a Galilean transformation of the form (\ref{NuoveCoord}), the horizontal velocity component transforms according to
$\dot{x}^* \= \dot{x} + u$ with arbitrary $u$, it follows that for each observer $\F^*$
the phase--space diagram must include a condition of the form
\begin{eqnarray}
	\dot{\vth} \sin\vth < k_{\F^*}(\dot{x}, u) \, .
\end{eqnarray}
Because of the arbitrariness of $u$, this implies that the exclusion of some of the critical points $P_{c1}^-,  P_{c2}^-, P_{c1}^+, P_{c2}^+$ depends on the choice of the inertial observer.

\begin{Remark}
	For completeness, we note that a subtle logical point may be raised regarding the ACM characterization as formulated in GB (p. 655, line 7) and recalled in (\ref{DryFrict}). While the implication $\vett{v} \ne 0 \, \Rightarrow \, F_T \, = \, - \, \mu \, |F_N| \frac{v}{|v|}$ is unquestionably valid, the  statement $\vett{v} = 0 \, \Rightarrow \, |F_T| \, \le \,\mu \, |F_N|$ need not hold. Indeed, the system dynamics may admit stopping instants of point $A$ at which the condition $|F_T| \, \le \,\mu \, |F_N|$, that is a condition determined by the forces acting on the system, is not instantaneously satisfied. Consequently, the system may move with a nonzero sliding velocity in $A$ immediately after the stopping instant without the condition $|F_T| \, \le \,\mu \, |F_N|$ being satisfied at any instant.
\end{Remark}

\section{Underlying causes of the critical issues}

The main reason for the problematic aspects highlighted in the previous section lies, as anticipated, in the choice of identifying the sliding velocity $v$ of the rod along the supporting line with the horizontal component $\dot{x}_A$ of the velocity (for the observer $\F$) of point $A$. If such a choice is adopted, then, by the Galilean velocity-addition theorem,
\begin{eqnarray}\label{TAV}
	\vett{v}^{\F}_A \= \vett{v}^{\F'}_A + \vett{u}
\end{eqnarray}
where $\F$ and $\F'$ denote two different observers, not necessarily inertial, and $\vett{u}$ is the transport velocity of $\F'$ relative to $\F$, the sliding velocity cannot be invariant with respect to the observer, not even in the case of inertial frames for which $\vett{u}$ is a constant vector.

It is well known (see \cite{Johnson}, p. 3) that the sliding velocity $\vett{v}^{(sl)}$
of a surface $\sigma$ over another surface $\sigma'$
is defined as the difference between the velocity  $\vett{v}_P$ of the point $P$ of $\sigma$ and the velocity $\vett{v}_{P'}$ of the point $P'$ of $\sigma'$, where $P$ and $P'$ are the points instantaneously in contact. Both velocities necessarily belong to the common tangent plane to the two surfaces at the contact point, since otherwise the motion would either produce an impact between the two surfaces or cause their mutual separation.
It is immediately apparent that, under this definition, the sliding velocity
\begin{eqnarray}\label{VelStrisc}
	\vett{v}^{(sl)} \= \vett{v}_P - \vett{v}_{P'}
\end{eqnarray}
is invariant with respect to the observer evaluating it, regardless of whether the latter is inertial.

In the specific case described in the previous section, the observers $\F$ and   $\F^*$
describe the contact constraint in the same geometric manner, namely through the condition
$y_A \= y - L \sin \vth = 0 $, but they describe it differently from a kinematic viewpoint: points of the constraint that are at rest according to observer $\F$ are not at rest according to observer $\F^*$, and vice versa.

Assuming that GB implicitly adopted, as the inertial frame for describing the CPP, the one in which the constraint is at rest, it should first be pointed out that, as will be shown later, the existence of such a frame should either be established or explicitly postulated. Moreover, if such a choice were indeed possible, then the equivalence between cases $M_{II}$ and $M_{III}$ claimed by GB up to a change of inertial frame, would no longer hold, since the inertial frame in which the constraint is at rest, if it exists, cannot depend on the direction of the sliding velocity.

\medskip

Note moreover that the invariance of the sliding velocity as correctly defined in (\ref{VelStrisc}) raises a further potential issue concerning the work of GB. Since the difference between two velocities is invariant under a change of observer, and since under suitable conditions the CPP exhibits an impact without collision that produces a velocity jump $\Delta \vett{v}_A \= \vett{v}^R_A - \vett{v}^L_A$, this jump should likewise be invariant with respect to the observer, possibly even beyond the class of inertial frames.

\medskip

We may therefore conclude this section by observing that, focusing on the invariance properties under changes of reference frame, GB's analysis involves then three distinct types of quantities: the ``absolute'' sliding velocity of the rod endpoint $A$, which is never invariant under a change of reference frame; the forces appearing in the equations of motion, which are invariant under changes of Galilean frame; and the velocity jump associated with the impact without collision, which is invariant under arbitrary changes of reference frame, including non-Galilean ones. These invariance properties of these three quantities do not appear to be sufficiently analysed in \cite{GenBro}.

\section{Kinematic Description of the Constraint}

In light of the observations made in the previous section, it is immediately apparent that the weaknesses identified in \cite{GenBro} are structurally related to the definition of a rough constraint, or, more generally, to the definition of a force that depends on velocity. For instance, the modelling of the resistive force acting on a mass point moving in a viscous fluid cannot disregard the motion of the fluid particles. Likewise, the characterization of the friction force exerted by an arbitrary rough constraint on a mechanical system (even in the simple case of a single mass point bilaterally constrained to move along a rough line), since it depends on the sliding velocity, cannot be based solely on a geometric description of the constraint, but also requires a kinematic description of the points belonging to the constraint itself.

Naturally, such a kinematic description is unnecessary whenever the characterization of the constraint depends only on its geometric properties, as in the case of a smooth constraint, whose reaction force is characterized simply by being orthogonal to the constraint.

A fortiori, the need for a kinematic description of the constraint applies to any analysis of the CPP based on Newton's second law and on the ACM characterization of the constraint. The use of Newton's second law entails the requirement of invariance under changes of inertial frame, whereas the ACM characterization requires the introduction of a kinematic description of the constraint. Incidentally, this observation remains valid even in the case where point $A$ is bilaterally constrained to the rough line, a circumstance that would eliminate the need to introduce the unilateral conditions (\ref{LCP}).

To illustrate how these considerations arise in the context of the CPP, let us consider the following three examples. Let $\widehat{\F}$ be a given inertial observer describing the upper Euclidean half-plane $\Euclideo^+_2$ by means of coordinates $(\xi, \eta)$ with $\eta \ge 0$. Temporarily disregarding the presence of the rod $AB$, we introduce the following possible kinematic descriptions of the boundary $\S$ of $\Euclideo^+_2$, each parameterized by the parameter $\lambda$.

\begin{description}
\item[Ex.1.] Let $\S_{\cal G}$  be described by
\begin{eqnarray}\label{VincoloGalileiano}
	\S_{\cal G} : \left\{
	\begin{array}{lcl}
		\xi(\lambda) & \= & \lambda + u_x t \\
		\eta(\lambda) &\= & 0
	\end{array}	\right.
\end{eqnarray}
where $u_x$ is a constant. We shall refer to this as a Galilean-type constraint.
\item[Ex.2.] Let  $\S_{\cal R}$ be described by
\begin{eqnarray}\label{VincoloArmonico}
	\S_{\cal R} : \left\{
	\begin{array}{lcl}
		\xi(\lambda) & \= & \lambda + A \cos(\omega t) \\
		\eta(\lambda) &\= & 0
	\end{array}	\right.
\end{eqnarray}
where $A,\omega$ are positive constants. We shall refer to this as a rigid-type constraint.
\item[Ex.3.] Let $\S_{\cal D}$  be described by
\begin{eqnarray}\label{VincoloElastico}
	\S_{\cal D} : \left\{
	\begin{array}{lcl}
		\xi(\lambda) & \= & \lambda + A \sin\left(\dfrac{\lambda}{B}\right)\cos(\omega t) \\
		\eta(\lambda) &\= & 0
	\end{array}	\right.
\end{eqnarray}
where $A<B,\omega$ are positive constants. We shall refer to this as a  deformable-type constraint.
\end{description}

All three kinematic descriptions have the same geometric support, namely the straight line $\eta=0$. However, in the three cases the points of this line undergo different motions when observed from the frame  $\widehat{\F}$. If one heuristically imagines the line $\eta=0$ as a conveyor belt, then in the first case its points move with uniform translational velocity; in the second case they undergo harmonic motion; and in the third case each point undergoes harmonic motion, but with an amplitude that depends on its distance from the origin.

More specifically, in the third case the points corresponding to $\lambda=k\pi B$ remain at rest, whereas the points corresponding to $\lambda=(\frac{\pi}{2}+k\pi )B$ undergo horizontal harmonic motion of amplitude $A$.

The rigid constraint condition can obviously be generalized by replacing the function $\cos(\omega t)$ with an arbitrary scalar function $f(t)$ depending only on time. Likewise, the deformable constraint condition can be generalized by replacing the term $ A\sin\left(\frac{\lambda}{B}\right)\cos(\omega t)$ with a scalar function $f(t,\lambda)$.

In the Galilean case, there exists an inertial frame $\F$ in which all points of the constraint are at rest. In the rigid case, no such inertial frame exists; however, there exists a non-inertial frame, rigidly attached to $\widehat{\F}$, in which the points of the constraint are at rest. Finally, in the deformable case, no frame rigidly attached to $\widehat{\F}$ can render all points of the constraint stationary.

Knowledge of the motion of the points of the constraint is indispensable for determining the sliding velocity of the endpoint $A$ of the rod relative to the constraint, which is in turn indispensable for determining the sense of the tangential component of the reaction force. Therefore, the kinematic description of the constraint must be regarded as part of the given data of the problem.

In the next section we shall briefly summarize which of the results obtained by GB can be immediately recovered within the framework of kinematically characterized constraints. We conclude this section, however, by observing that analogous issues concerning the invariance of the results under changes of inertial observer may arise in any analysis of a mechanical system based on quantities that depend explicitly on velocity. A notable example is provided by collision theory, where analyses based on the balance of the system's kinetic energy may exhibit similar difficulties.

\section{Reappraisal of GB and alternative approaches}

A possible way to improve the analysis of the CPP presented by GB is therefore to adapt, whenever possible, the GB approach to the framework arising from the introduction of the kinematic aspects of the constraint.

In the case of a Galilean constraint, the difference between the analysis presented by GB and the adapted formulation is essentially conceptual. It consists in selecting, for the description of the CPP, the specific inertial observer $\F$ for which the points of the constraint are at rest. In such a frame $\F$, the following properties hold:

\begin{enumerate}
\item the sliding velocity $\vett{v} \= v \, \vett{e}_x$ is indeed given by $v \= \dot{x}_A \= \dot{x} + L \dot{\vth} \sin \vth$;

\item every other inertial observer measures the same sliding velocity; consequently, the condition $\dot{x}_A \= \dot{x} + L \dot{\vth} \sin \vth <0 $ is invariant under a change of inertial observer;

\item the analogy between mode $\M_{II}$, characterized by $\dot{x}_A<0$ and $F_T >0$, and mode $\M_{III}$, characterized by $\dot{x}_A>0$ and $F_T < 0$, remains acceptable. However, it can no longer be justified by a change of inertial observer. Rather, it follows from the symmetry of the function $\B(\vth,\mu) \= \dfrac{1}{m} \left(1+3\cos\vth(\cos\vth - \mu \sin\vth)\right)$ with respect to the axis $\vth = \frac{\pi}{2}$, or equivalently from the identities $\sin(\pi - \vth) = \sin\vth, \,\, \cos(\pi - \vth) = - \cos\vth$;

\item the condition $\dot{\vth} \sin \vth < - \dfrac{\dot{x}}{L}$,
which is now invariant under changes of Galilean observer, must nevertheless be taken into account in the analysis of the phase portrait shown in (\ref{Fig2}). As a consequence, some critical points associated with possible motions of the system become unreachable. This effectively yields an improvement of the qualitative analysis, showing that the behaviour of the CPP depends also on the horizontal component of the velocity of the rod's centre of mass.
\end{enumerate}
Therefore, once the purely formal aspects of the approach are set aside, point (iv) emerges as the only aspect that represents a genuinely significant improvement over the analysis of GB. Nevertheless, the existence of a frame of rest for the constraint remains an important feature of the CPP, particularly since the other two possible cases, namely rigid and deformable constraints, can both be naturally encompassed within the kinematic description of the constraint. Moreover, the issue of the invariance, under a change of reference frame, of the velocity jump associated with the impact without collision derived by GB in their analysis still remains to be investigated.

\medskip

In the case of a rigid but non-Galilean constraint, the conceptual difference between the original GB formulation and the adapted one again lies in the mandatory choice of the frame $\F$ in which the constraint is instantaneously at rest. In this case, however, the equations of motion for the CPP must also include the apparent forces. These forces are regarded as known quantities, since the motion of the constraint—and therefore the motion of $\F$ relative to an inertial frame—is assumed to be known a priori.

Finally, in the case of a deformable constraint, no rigid observer can see all points of the constraint at rest. The determination of the sliding velocity at a given instant is possible only if the instantaneous motion state of the rod is known at that instant. This suggests approaching the CPP in an event-driven manner: once the motion state of the rod $AB$ is known at the initial time $t_0$, one determines an evolution of the system consistent with the laws of Dynamics.

Such an evolution may be governed either by the ordinary form of Newton's second law, thereby determining the motion of the system in a right neighbourhood of $t_0$, or by Newton's second law in impulsive form,
$\vett{P} \= \Delta \vett{v}$, thereby generating a new set of initial conditions for the CPP, compatible with the constraint and from which the system subsequently evolves according to the ordinary form of Newton's second law.

\medskip

As already observed, however, the use of Newton's second law in impulsive form would require the evolution to be invariant with respect to any observer, including non-Galilean ones.

In such a setting, one may therefore consider the more radical alternative of studying the CPP by focusing exclusively on its impulsive aspects. This entails abandoning the objective of describing the continuous motion of the rod while in contact with the rough line, and instead analysing only the instantaneous behaviour of the rod at contact as a function of its current motion state. At the same time, the ordinary form of Newton's second law is replaced by its impulsive counterpart.

Such an approach can be developed within a fully frame-invariant framework (see \cite{PasqueroPainleve-Arxiv}) and eliminates, at their root, the difficulties identified here in connection with \cite{GenBro}. However, as already emphasized, it provides only local, instantaneous information and not a global description of the motion of the rod.

\section*{Acknowledgements and declarations}

The author acknowledges the support of University of Parma and of the Italian National Group of Mathematical Physics (GNFM-INdAM).
%

During the preparation of this manuscript, the author used an artificial intelligence-based language assistance tool (ChatGPT) exclusively to improve the grammar, spelling, readability, and overall clarity of the text. The AI tool was not used to generate scientific content, perform data analysis, interpret results, formulate conclusions, or make scientific decisions. All scientific content, analyses, interpretations, and conclusions presented in this manuscript are the sole responsibility of the author, who carefully reviewed and edited the manuscript. 

\end{document}